\documentclass[runningheads]{llncs}
\usepackage{graphicx}
\usepackage{subcaption}
\usepackage{float}
\begin{document}
\title{GuessTheMusic: Song Identification from
Electroencephalography response\thanks{Supported by organization x.}}
%
%
\author{Dhananjay Sonawane\inst{1} \and
Krishna Prasad Miyapuram\inst{2} \and
Bharatesh RS\inst{2} \and
Derek J. Lomas\inst{3}}
\authorrunning{D. Sonawane et al.}

%
\institute{Computer Science and Engineering, Indian Institute of Technology Gandhinagar, Gujarat - 382355, India
\email{dhananjay.sonawane@alumni@iitgn.ac.in} \and
Centre for Cognitive and Brain Sciences, Indian Institute of Technology Gandhinagar, Gujarat - 382355, India
\email{kprasad@iitgn.ac.in}\\ \and
Industrial Design Engineering, Delft University of Technology, Netherlands
\email{J.D.Lomas@tudelft.nl}\\}
\maketitle              
\begin{abstract}
The music signal comprises of different features like rhythm, timbre, melody, harmony.
Its impact on the human brain has been an exciting research topic for the past several decades. Electroencephalography (EEG) signal enables non-invasive measurement of brain activity. Leveraging the recent advancements in deep learning, we proposed a novel approach for song identification using a Convolution Neural network given the electroencephalography (EEG) responses. We recorded the EEG signals from a group of 20 participants while listening to a set of 12 song clips, each of approximately 2 minutes, that were presented in random order. The repeating nature of Music is captured by a data slicing approach considering brain signals of 1 second duration as representative of each song clip. More specifically, we predict the song corresponding to one second of EEG data when given as input rather than a complete two-minute response. We have also discussed pre-processing steps to handle large dimensions of a dataset and various CNN architectures. For all the experiments, we have considered each participant's EEG response for each song in both train and test data. 
We have obtained 84.96\%  accuracy for the same. The performance observed gives appropriate implication towards the notion that listening to a song creates specific patterns in the brain, and these patterns vary from person to person. 

\keywords{EEG \and CNN \and neural entrainmment \and music \and frequency following response  \and brain signals \and classification}
\end{abstract}
\section{Introduction}
Audio is a type of time-series signal characterized by frequency and amplitude.
Music signals are a particular type of audio signals that posses a specific type of acoustic and structural features.
Accordingly, one would expect that music affects different parts of the brain as compared to other audio signals. Nevertheless, how closely the brain activity pattern is related to the perception of periodic signal like music? Electroencephalography (EEG) is a method to measure electrical activity generated by the synchronized activity of neurons. There is plethora of evidence published by researchers to bolster linkage between EEG response and music. 
Braticco et al.\cite{b0} showed that the brain anticipates the melodic information before the onset of the stimulus and that they are processed in the secondary auditory cortex. 
The study on the processing of the rhythms by Snyder et al. found out that the gamma activity in the EEG response corresponds to the beats in the simple rhythms\cite{b0_}.
A recent study showed that it is possible to extract tempo - a critical music stimuli feature, from the EEG signal\cite{b1}. The authors concluded that the quality of tempo estimation was highly dependent on the music stimulus used. 
The frequency of neural response generated after entrainment of music is highly related to its beat frequency\cite{b2}.
Further, few researchers carried out Canonical Correlation Analysis to estimate the correlation coefficient of music stimuli with EEG data\cite{b3,b4}.

However, work done on pattern of brain activity reflecting neural entrainment to music listening and its recognition is still at an early stage. These patterns are very much intricated and thus it is hard to interpret what is happening in the human brain when a person is listening to a song. Moreover, aesthetic experience associated music listening is highly subjective - i.e. it varies from person to person and also from time to time depending on various contextual factors such as mood of the individual who is listening to music. That is why the song identification task is challenging.
Previous research has focused on the relationship between the song and its brain (EEG) responses. They have used engineered features for processing the EEG data which are dependent on the domain knowledge. There have been few attempts on automatic feature extraction from EEG data using neural networks for song classification task\cite{b5}.

Taking the notion of the resonance between EEG signal and music stimuli, in this paper, we hypothesize the following - 1) music stimuli create identifiable patterns in EEG response 2) for a given song; these patterns vary from person to person. 
We pose these hypothesis as a song identification task using deep learning architecture. 
To study the first hypothesis, we split each participant's EEG response for each song in training and test dataset. We explored how large the train data should be and the effect of train data size on the performance of the model.
For a given participant, the model learns song pattern present in EEG response from training data and try to predict song ID for test data.
For the second hypothesis, we exclude some participants entirely from the training dataset. 
During data preprocessing, raw EEG response is divided into segments, and each such chunk corresponds to 1 second long EEG response.
Our model predicts song ID for each such chunk present in test data. It is represented as 2D matrices, and we call them "song image". More about data preprocessing is discussed in section 3.B.
This technique allows us to use 2D and 3D convolution neural networks, which are usually used in computer vision and image processing fields.
The features extracted from the song image by CNN are fed to a multilayer perceptron network for the classification task. Our results outperform state of the art accuracy. 

The remaining part of the paper is organized as follows: section 2 describes prior work on song classification problem using EEG data and their results.
Section 3 reports our methods, including data collection, pre-processing steps, and CNN architectures used.
In section 4, we discuss the performance of our model, and in section 5, we draw a conclusion on the cognitive process behind music perception and suggest possible future work.  

\section{Related work}
It has been shown that human mental states can be unraveled from non-invasive measurements of brain activity such as functional MRI, EEG etc \cite{b6}. 
Several researchers have documented the frequency following response (FFR), which is the potential induced in the brain while listening to periodic or nearly periodic audio signal\cite{b7}.
A successful attempt has been made to reconstruct perceptual input from EEG response. In \cite{b8}, the objective of the study was: a person is looking at an image, and brain activity is captured by the EEG in real-time. The EEG signals are then analyzed and transformed into a semantically similar image to the one that the person is looking at. They modeled the Brain2Image system using variational autoencoders (VAE) and generative adversarial networks (GAN). 
The research work\cite{b9} relates music and its activity using a statistical framework. Authors study the classification of musical content via the individual EEG responses by targeting three tasks: stimuli-specific classification, group classification, i.e., songs recorded with lyrics, songs recorded without lyrics, and instrumental pieces and meter classification, i.e., 3/4 vs. 4/4 meter. They have used OpenMIIR dataset\cite{b10}. It includes response data of 10 subjects who listened to 12 music fragments with duration ranging from 7 s to 16 s coming from popular musical pieces. They proposed Hidden Markov Model and probabilistic computation method to the developed model, which was trained on 9 subjects and tested on the 10\textsuperscript{th} subject. They achieved 42.7\%, 49.6\%, 68.7\% classification rate for task1, task2, task3, respectively. 
Foster et al. investigated the correlation between EEG response and music features extracted by the librosa library in Python\cite{b11}. Using representational similarity analysis, they report the correlation coefficient of EEG data with normalized tempogram features as 0.63 and with MFCCs as 0.62. They also deal with song identification from the EEG data problem and obtained 28.7\% accuracy using the logistic regression model.
Our study stands out in terms of methodology as we exploit the power of deep learning architecture for automatic feature extraction from EEG response.

Yi Yu et al. used a convolution neural network called DenseNet\cite{densenet} for audio event classification\cite{b5}. The EEG responses were collected on 9 male participants. The audio stimuli were 10 seconds long, spanned over 8 different categories (Chant, Child singing, Choir, Female singing, Male singing, Rapping, Synthetic singing, and Yodeling). They achieved 69\% accuracy using EEG data only. However, the optimal result was 81\%, where they used audio features extracted from another convolution network - VGG16\cite{vgg} along with EEG response.
Sebastian Stober et al., aimed to classify 24 music stimuli\cite{b13}. Each music segment comprised of 2 unique rhythms played at a different pitch. Due to small data, they process and classify each EEG channel individually. CNN was trained on 84\% of complete response (approximately 27 second response chunk out of 32 seconds), validated on 8\% of complete response(approximately 2.5 second response chunk), and tested on 8\% of complete response(approximately 2.5 second response chunk). They report 24.4\% accuracy. We, in this study, deal with more complex data. Our music segment comprises diverse tone, rhythm, pitch, and some of them also include vocals. This makes the song identification task more challenging. The proposed architecture is much simpler compared to DenseNet\cite{densenet} and VGG16\cite{vgg}.

\section{Methods}
The aim of this work is to create an approach to classify EEG response corresponding to respective song event. The complete experiment can be described in 4 phases : Data collection, Preprocessing, CNN architecture and model development, Testing models.
\subsection{Data Collection}
Participants were made to sit in a dimly lit room. Then we collected demographic information like age, gender, and handedness. 
Brief information regarding the EEG collection setup, time that the experiment will take, the responses that they have to make were discussed with all participants. 
Then we measure the circumference of the participant's head to select a suitable EEG cap.
\begin{figure}
    \begin{center}
        \includegraphics[width=0.45\textwidth]{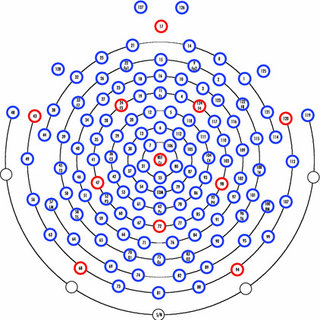}
    \end{center}
\caption{Illustration of the 128-channel-system and electrode position}
\label{ele_pos}
\end{figure}
The 128 channel high density Geodesic electrode net cap (Hydrocel Geodesic SensorNet platform, Electrical Geodesics Inc., USA, Now Philips) is chosen according to the headsize measurement.
The cap is immersed in the KCl electrolyte solution prepared in 1 litre pure distilled water.
The reference electrode position is measured as the intersecting point on the lines between nasion (point in between eyebrows) and inion (middle point of skull ending at the backside) with preauricular points on both sides, and then it is marked.
Other electrodes are placed according to the International 10-20 system, as depicted in Fig.\ref{ele_pos}.

After this setup, participants were asked to close the eyes on a single beep tone. 
This is followed by 10 seconds of silence, after which the song stimulus is presented.
At the end of each stimulus, a double beep tone was sounded at which the participants were instructed to open eyes and make a response.
They were asked two questions :
\begin{itemize}
    \item How much familiar are you with the song? The participants were asked to rate in the range of 1 to 5(where one indicated strongly familiar, and five indicated strongly unfamiliar)
    \item How much did you enjoy the song? The participants were asked to rate in the 33 range of 1 to 5(where one indicated extremely enjoyable, and five indicated incredibly dull)
\end{itemize}

\noindent These responses were collected within 10 second silence window before presenting the next song. 
Since the maximum length of the song is 132 seconds, all other responses were zero padded accordingly. 
Therefore, all song responses are of 142 seconds after considering the above window. 
Total of 20 participants data collected on 12 music stimuli.
\begin{table}
\caption{Songs used in EEG data collection}
\begin{center}
\begin{tabular}{|c|c|c|c|}
\hline
\textbf{Song ID}& \textbf{Song Name}& \textbf{Artist}& \textbf{Song Length} \\
\textbf{}& \textbf{}& \textbf{}& \textbf{(in seconds)} \\
\hline
1& Trip to the lonely planet& Mark Alow& 125  \\
\hline
2& Sail& Awolnation& 114  \\
\hline
3& Concept 15& Kodomo& 132  \\
\hline
4& Aurore& Claire David& 111  \\
\hline
5& Proof& Idiotape& 124  \\
\hline
6& Glider& Tycho & 100  \\
\hline
7& Raga Behag& Pandit Hari Prasad Chaurasiya& 116  \\
\hline
8& Albela sajan& Shankar Mahadevan& 121  \\
\hline
9& Mor Bani Thanghat Kare& Aditi Paul& 126  \\
\hline
10& Red Suit& DJ David. G& 129  \\
\hline
11& Fly Kicks& DJ Kimera& 113  \\
\hline
12& JB& Nobody.one& 117  \\
\hline
\end{tabular}
\label{song_list}
\end{center}
\end{table}
Songs used in the experiment are listed in Table~\ref{song_list}.
The songs contain some tonal and vocal excerpts.
The sampling rate for 11 participants was 1000Hz, and for the remaining 9 participants, it was 250Hz. 
16 participants were male, while 4 were female. All of them right-handed with an average age of 25.3 years and a standard deviation of 3.38. 
All music stimuli were presented to the subject in random order. 

\subsection{Data Preprocessing}
EEG is highly sensitive to noise. It also captures eye blinking and high-frequency muscle movements. 
Therefore, it is necessary to clean EEG data before using it for any application. 
EEGLAB toolbox was used to implement the majority of preprocessing steps. 
Once the channel locations are provided, we performed average re-referencing with respect to channel number 129. 
Raw EEG signal was then loaded as epochs for each presented song. 
12 epochs were created for each participant. 
By this, we eliminated the signals which do not lie in our area of interest.
We then used independent component analysis to remove artifact data. 
This has been achieved using 'runica' algorithm present in MATLAB. We have used 'adjust' toolbox for artifact removal.
For simplicity, positive infinities, negative infinities, and NaN (not a number) values are replaced by zero. 
However, taking the average value of surrounding electrodes for these outliers would be a better approach and may improve the performance.
The above steps create final ready-to-use data.

For deep learning models, we would need extensive data with reasonable feature dimensions to detect a pattern. 
In our task, we had the opposite situation. Our data contains 240 EEG responses corresponding to 20 participants and 12 song stimuli. 
However, a number of samples collected in one electrode for one song of one participant is greater than 27,000 at the 250Hz sampling rate.
The number of samples goes beyond 100,000 at 1000Hz sampling rate. Thus, our data consists of a few examples with very high dimensionality.
\begin{figure}
    \centering
    \begin{subfigure}[b]{0.3\textwidth}
        \includegraphics[width=5cm, height =4cm]{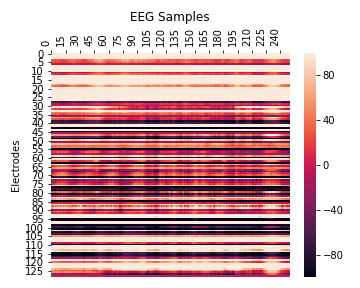}
       \caption{Song ID - 6}
        \label{time_example_song_6}
    \end{subfigure}
    ~ 
    \qquad
    \begin{subfigure}[b]{0.3\textwidth}
        \includegraphics[width=5cm, height =4cm]{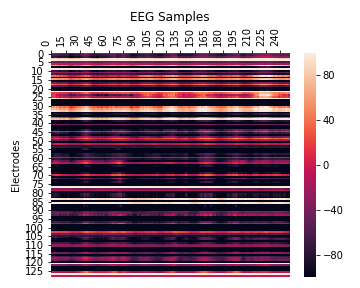}
        \caption{Song ID - 7}
        \label{time_example_song_7}
    \end{subfigure}
    \caption{Song images of 26\textsuperscript{th} second for participant ID: 1902 in time domain}
    \label{time_song_image}
\end{figure}

\begin{figure}
\begin{center}
    \includegraphics[width=0.65\textwidth]{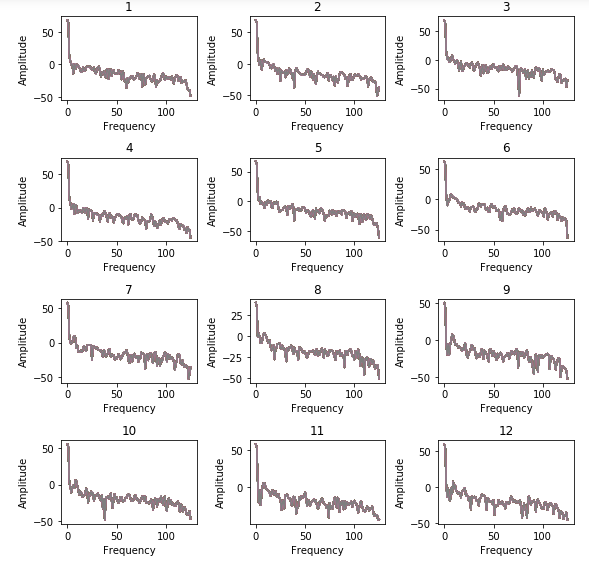}
\end{center}
\caption{FFT of EEG response of the participant ID 1902 for all 12 songs}
\label{fft}
\end{figure}

The data augmentation technique was used to increase data. 
We split the EEG response of each participant for each song in a chunk of 1-second long window. 
Thus, giving us 128(electrodes) * 250 or 1000 (samples per second) dimension 2D matrices. 
We call them "song image" and label them the same song ID of the corresponding song. 
Such a formulation not only increases number of examples in original dataset but also allows us to use 2D, and 3D convolution networks.
Fig.~\ref{time_song_image} shows the song images of the 26\textsuperscript{th} second of two different songs for one of the participants in time domain. 
However, window size is design parameter and it is difficult to decide optimal window size for obtaining song images.
Larger window size increases dimensionality of a song image and decreases the total number of song examples. Thereby killing the purpose of the data augmentation.
High dimensional input to Convolution Neural Network (CNN) also drastically increases number of trainable parameters provided that the rest of the architecture remains the same.
Smaller window size will carry less information about EEG signal and CNN may perform poorly.
As the sampling rate changes, the 1 second time window will carry different number of samples in song image. This causes inconsistency in input data to the CNN.
We needed more concrete preprocessing steps so that smaller window size would increase number of examples in dataset but not at the cost of the performance.
Fig.~\ref{fft} illustrates the Fast Fourier Transform (FFT) of the all 12 song response for one participant. 
It is worth noting that, in all the FFT's, the maximum frequency component is less than 100Hz.
This is expected as EEG is characterized by frequency bands -  0Hz - 4Hz (Delta band), 4Hz - 8Hz (Theta band), 8Hz - 15Hz (Alpha band), 15Hz - 32Hz (Beta band), and frequencies higher than 32Hz (Gamma band), exhibit high power in low frequency ranges.

Using spectopo function in EEGLAB, we converted time-domain EEG data into the frequency domain.
Spectopo calculates the amplitude of a frequency component present in each 1 second window of the EEG response.
The maximum frequency component in the frequency domain representation of data is chosen as per Nyquist's criteria. 
It is 125Hz and 500Hz when the sampling rate is 250Hz and 1000Hz, respectively.
Regardless of what is sampling frequency of EEG, we can safely choose 125Hz as maximum frequency component. 
\begin{figure}
\begin{center}
    \includegraphics[width=0.6\textwidth]{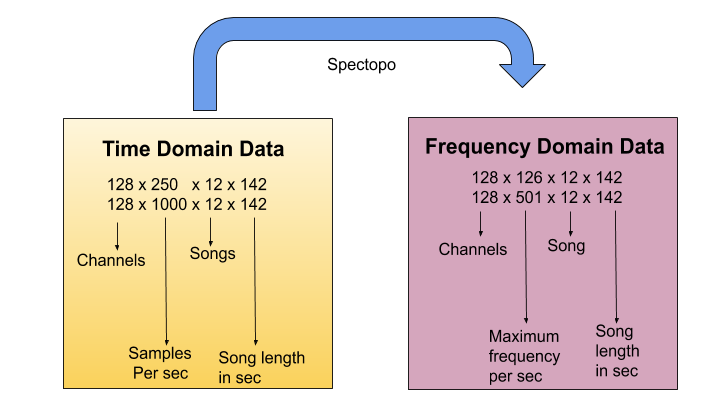}
\end{center}
\caption{Time domain to frequency domain conversion for one participant data using spectopo}
\label{time_to_freq}
\end{figure}
Fig.~\ref{time_to_freq} explains the dimensionality and conversion of time-domain data to frequency domain data. 
Frequency domain data helps in dimensionality reduction as well as makes input dataset consistent and compact. 
But, the time window for which we calculate FFT is again a design parameter.
The effect of the time window on a performance of the CNN in both time and frequency domain has been further addressed in next section.

\begin{figure}
    \centering
    \begin{subfigure}[b]{0.3\textwidth}
        \includegraphics[width=5cm, height =4cm]{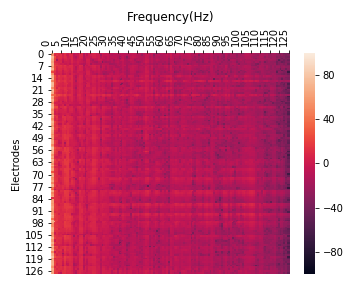}
       \caption{Song ID - 6}
        \label{freq_example_song_6}
    \end{subfigure}
    ~ 
    \qquad
    \begin{subfigure}[b]{0.3\textwidth}
        \includegraphics[width=5cm, height =4cm]{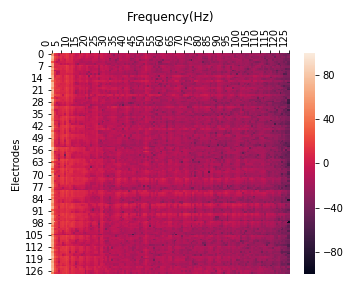}
        \caption{Song ID - 7}
        \label{freq_example_song_7}
    \end{subfigure}
    \caption{Song images of 26\textsuperscript{th} second for participant ID: 1902 in frequency domain}
    \label{freq_song_image}
\end{figure}
\noindent The row of song image in frequency domain denotes electrode while column denotes the frequency.
Fig.~\ref{freq_song_image} shows the song images of the 26\textsuperscript{th} second of the different song for participant 1902 in frequency domain.

We follow standard practice in machine learning to develop the model. 
Test-Train split is 0.3 with a random selection of song images in training (70\%) and testing  (30\%) data.
Validation split parameter is set to 0.2.

\subsection{Experiment and Model development}

Convolutional Neural Network is the core part of our model because it learns the underlying pattern in the song image. 
It includes many hyperparameters that need to be set carefully.
We apply CNN to the both time domain and frequency domain song image dataset.
The CNN architecture remains same except for the input layer where shape of the input song image changes as per domain and time window.

\begin{figure*}
\centerline{\includegraphics[width=1.0\textwidth]{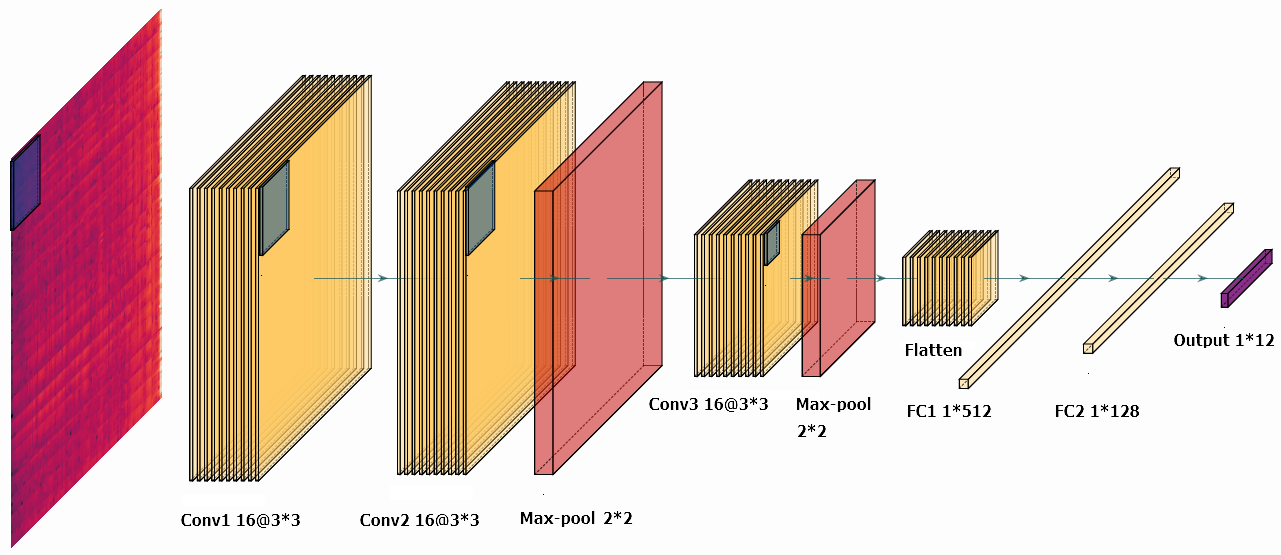}}
\caption{2D CNN architecture followed by Dense network for song classification}
\label{2d_cnn}
\end{figure*}

To study our first hypothesis that music creates an identifiable pattern in the brain EEG signals, we consider all participants data in training data. 
An immediate problem is how much of 1 participant's response should be considered in training data to predict song ID? 
To answer this question, we vary the train-test split from 20\% to 95\%. 
The $x$\% train-test split means $x$\% of data will be chosen as test data while $(100-x)$\% will be treated as train data. 
For each split, we randomly select test samples. We have created a 3-layer CNN network for feature extraction and a 2-layer dense neural network for song classification. 
Except for the last output layer, each convolution layer as well as the dense layer has the ReLu\cite{relu} activation function. 
It brings non-linearity in architecture and helps to detect complex patterns. 
Since we are doing multiclass classification, the output layer has a softmax activation function. 
The loss function used is categorical cross-entropy. 
It is given by,
\begin{equation}
    Cross entropy = - \sum_{i=1}^{C} y_{o,c} * log(p_{o,c}),\label{cross_entr}
\end{equation}
\noindent where, $C$ is total number of classes - 12 in our case, $y_{o,c}$ is a binary number if class label $c$ is correct classification for observation $o$, $p_{o,c}$ predicted probability observation $o$ is of class $c$.

We used Adam optimizer to minimize categorical cross-entropy loss.
The kernel size is 3*3, and we have used 16 such filters at each convolution layer. 
Two max-pooling layers have been added after convolution layers 2 and 3. 
Their exclusion almost doubles total trainable parameter, thereby increasing network complexity, training time. 
Fig.\ref{2d_cnn} shows the 2D CNN architecture. Removing one or more layers from the architecture mentioned above resulted in an underdetermined system and failed to learn all the patterns. 
Adding extra layers led to overfitting and thus reducing the performance. 

We have analyzed the effect of the time window on the performance of the CNN in time domain. 
We created a 3 separate datasets with time window set to 1 second, 2 seconds, 3 seconds and having the song image shape 128*250, 128*500, 128*750 respectively. 
We did not increase time window beyond 3 seconds as number of examples in dataset were reduced significantly.
The 9 participants with EEG sampling frequency of 250Hz were chosen in above dataset. 
Similar steps were applied to the participant where EEG sampling rate was 1000Hz. To maintain the symmetry, we randomly chose 9 participants out of 11.
In frequency domain, we did not increase the time window for which FFT is calculated. Because, 1 sec time window gave reasonable results.
We have also analyzed the effect of higher sampling frequency. 
For this, we choose those 11 participants for which sampling rate was 1000Hz. This resulted in the maximum frequency component being 500Hz, thus song image shape changes to 128*501. The previous model was trained on new data. We could have compared this result with an earlier model where the song image was 128*126, but since earlier data had 20 participants. For a fair comparison, we choose the same 11 participants and discard all the frequency from 127Hz to 501Hz. We trained the model on this data as well.

For the investigation of our second hypothesis - music creates a different pattern on a different person; we used 5 of the randomly chosen participant's responses as a test dataset.
Training data included the remaining 15 participant's responses.
To improve the result for this task, we developed a 3D CNN model.
All the parameters remain same as the previous model except for the kernel, which changed to 3*3*3, and the max-pool layer modified as 2*2*2. 
We stacked 10 consecutive song images and fed it to the first layer of 3D CNN as input. 
The choice 10 is made by considering the trade-off between the number of samples in data and dimensionality of each 3D input sample.

We used 30 epochs for training the CNN. 
The network architecture which we have proposed is designed in the Keras framework. 
The initialization of weights is assigned with some random numbers by using the Keras framework\cite{keras} itself. 
The GPU NVIDIA GTX 1050 that we have used for this experiment has 4GB of RAM. 
The batch size was kept 16 for all the experiments because of memory constraints. 

\section{Result and discussion}

\begin{figure}
    \centering
    \begin{subfigure}[t]{0.4\textwidth}
        \includegraphics[width=6cm,height=4cm]{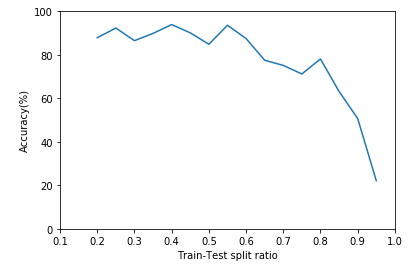}
       \caption{Change in the test accuracy for different train-test split values}
        \label{accu_train}
    \end{subfigure}
    ~ 
    \hspace{5mm}
    \begin{subfigure}[t]{0.3\textwidth}
        \includegraphics[width=5cm,height=4cm]{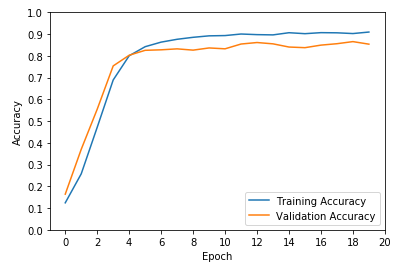}
        \caption{Training and validation curve}
        \label{training_curve}
    \end{subfigure}
    \caption{Accuracy plots}
    \label{training_plots}
\end{figure}

\begin{figure}
    \centering
    \begin{subfigure}[b]{0.3\textwidth}
        \includegraphics[width=\textwidth]{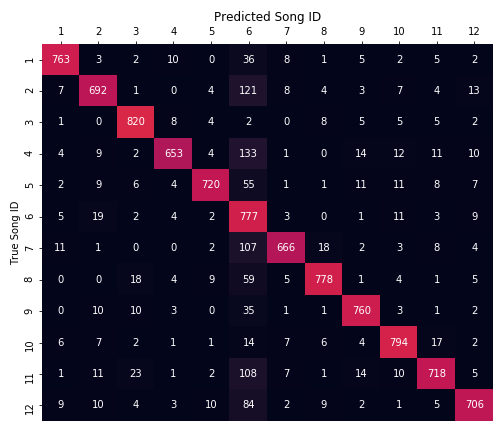}
        \caption{Confusion matrix for 0.3 train-test split}
        \label{conf_70-30}
    \end{subfigure}
    ~ 
    \begin{subfigure}[b]{0.3\textwidth}
        \includegraphics[width=\textwidth]{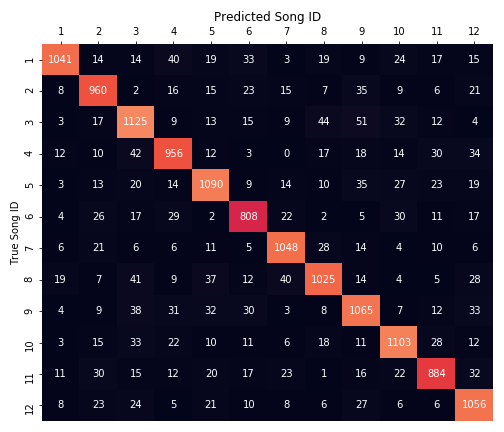}
        \caption{Confusion matrix for 0.5 train-test split}
        \label{conf_50-50}
    \end{subfigure}
    \begin{subfigure}[b]{0.3\textwidth}
        \includegraphics[width=\textwidth]{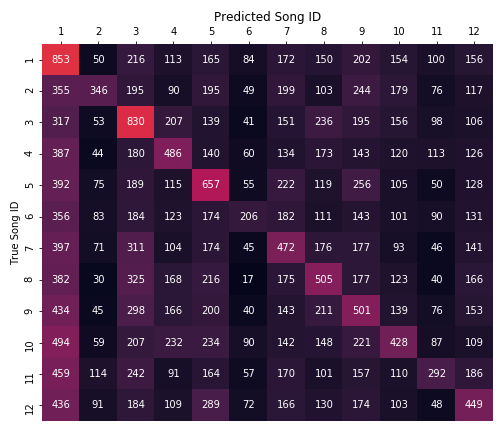}
        \caption{Confusion matrix for 0.95 train-test split}
        \label{conf_5-95}
    \end{subfigure}
    \caption{Confusion matrices}
    \label{all_cf}
\end{figure}

When each participant's response is accounted for in both train and test data, our model reports outstanding training accuracy as 90.90\% and test accuracy as 84.96\% in frequency domain (model 5). 
Confusion matrix is shown in Fig.~\ref{conf_70-30}. It shows that test data is well spread across all 12 classes, and almost all of them are correctly classified. 
Fig.~\ref{training_curve} shows accuracy vs. epoch curve for the above model.
Table~\ref{accu_table} summarizes the accuracy of the all CNN models.
\begin{table}
\caption{Performance of the CNN models}
\begin{center}
\begin{tabular}{|c|c|c|c|c|c|c|}
\hline
\textbf{CNN}& \textbf{Domain}& \textbf{Song image}& \textbf{Total number}& \textbf{CNN}& \textbf{Train}& \textbf{Test} \\
\textbf{Model}& \textbf{}& \textbf{shape}& \textbf{of song}& \textbf{trainable}& \textbf{accuracy}& \textbf{accuracy} \\
\textbf{}& \textbf{}& \textbf{}& \textbf{images}& \textbf{parameters}& \textbf{(\%)}& \textbf{(\%)} \\
\hline
1& Time& 128*250& 11,772& 3,398,476& 8.60& 8.01  \\
\hline
2& Time& 128*500& 5,832& 6,953,804& 8.64& 7.48  \\
\hline
3& Time& 128*750& 3,888& 10,623,820& 8.09& 8.82 \\
\hline
4& Time& 128*1000& 11,772& 14,179,148& 8.65& 8.05  \\
\hline
5& Time& 128*2000& 5,832& 28,515,148& 8.79& 7.45  \\
\hline
6& Time& 128*3000& 3,888& 42,851,148& 8.48& 8.29 \\
\hline
7& Frequency& 128*126& 34,080& 1,678,156& 90.90& 84.96\\
\hline
8& Frequency& 128*500& 18,774& 5,823,308& 80.01& 80.99\\
\hline
9& Frequency& 128*126& 18,774& 1,678,156& 91.31& 76.19\\
\hline
10& Frequency& 128*126& 34,080& 1,678,156& 86.95& 7.73\\
\textbf{}& \textbf{}& \textbf{}& Cross participant& \textbf{}& \textbf{}& \textbf{} \\
\hline
11& Frequency& 128*126*10& 3600& 1,925,228& 9.44&9.44\\
\textbf{}& \textbf{}& 3D input& Cross participant& \textbf{}& \textbf{}& \textbf{} \\
\hline

\end{tabular}
\label{accu_table}
\end{center}
\end{table}
The model trained on time domain dataset hardly learnt anything for song identification task. 
By changing the time window and sampling frequency did not change the performance of the CNN in time domain.
But the same CNN architecture obtained high accuracy when trained on frequency domain dataset. 
The performance of CNN in frequency domain could be either due to learning the temporal pattern in EEG or learning from other participant's responses.
To examine the latter cause, we retrained the same CNN model, this time excluding 5 participants entirely from the train data. We got 86.95\% training accuracy, but the model reported 7.73\% test accuracy (model 10). We extend this experiment by training the 3D CNN model, and observed 9.44\% test accuracy for  cross-participant data (model 11).
This shows that CNN depends upon the temporal features in each participant's response for the song prediction.
It also gives insights regarding the EEG pattern generated due to music entrainment differ from person to person for the same song. 

We also studied the high-frequency signals generated in the brain due to music entrainment.
For this, we choose those participants whose data is collected at 1000Hz. It will ensure the high value of the maximum frequency component(up to 500Hz) in the song image. All participant responses were considered in train and test data.
Two models of the same architecture were developed; one trained on data having all 500Hz frequency element while the other trained on data by choosing only the first 126Hz frequencies out of 500Hz. Both performed almost equally well, giving 80.99\% and 76.19\% accuracy, respectively (model 8,9).
It explains that higher EEG frequencies do not contribute much to the pattern generated while listening to music. To investigate how much each participant's data should be included in train data to predict song ID on test data, we vary the train-test split from 20\% to 95\%. Fig. \ref{accu_train} shows accuracy plot for different train-test split value.
We got remarkable test accuracy 78.12\% by training the CNN model on 20\% of total data(Train-test split = 0.8). In other words, by learning approximately 17 seconds long EEG response of 120 seconds prolonged music stimuli, we were able to predict song ID for the rest of the 103 seconds with 78\% correct prediction probability.
More commendable accuracy is 22.12\% at 0.95 train-test split. This performance is much better than a random guess, which is 8.33\% for this 12 class classification problem.
Fig.\ref{conf_70-30}, \ref{conf_50-50}, \ref{conf_5-95} shows the confusion matrices for 0.3, 0.5, 0.95 train-test split ratio, respectively. 

\begin{figure}
    \centering
    \begin{subfigure}[t]{0.45\textwidth}
        \includegraphics[width=5.4cm,height=5.4cm]{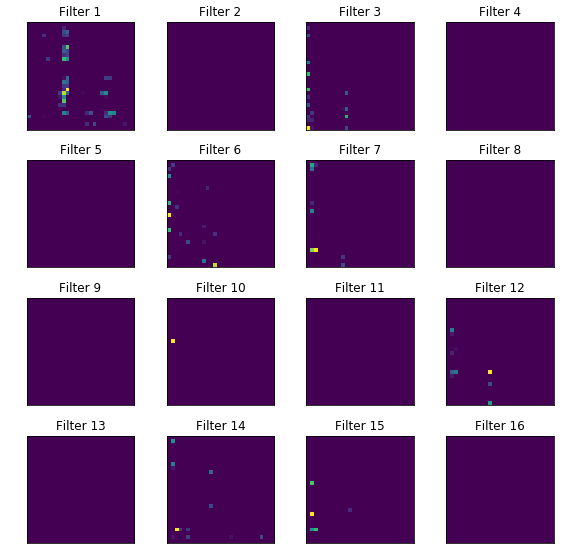}
        \caption{Participant ID: 1901, Song ID:9}
        \label{1901_9}
    \end{subfigure}
    ~ 
    \hspace{2mm}
    \begin{subfigure}[t]{0.45\textwidth}
        \includegraphics[width=5.4cm,height=5.4cm]{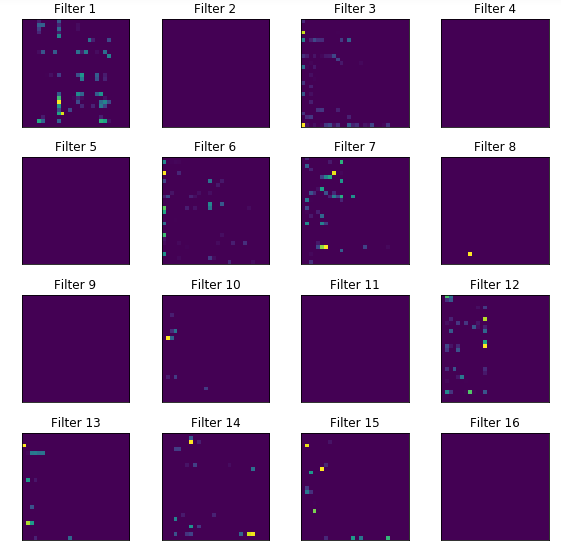}
        \caption{Participant ID: 1905, Song ID:9}
        \label{1905_9}
    \end{subfigure}
    
    \begin{subfigure}[b]{0.45\textwidth}
        \includegraphics[width=\textwidth]{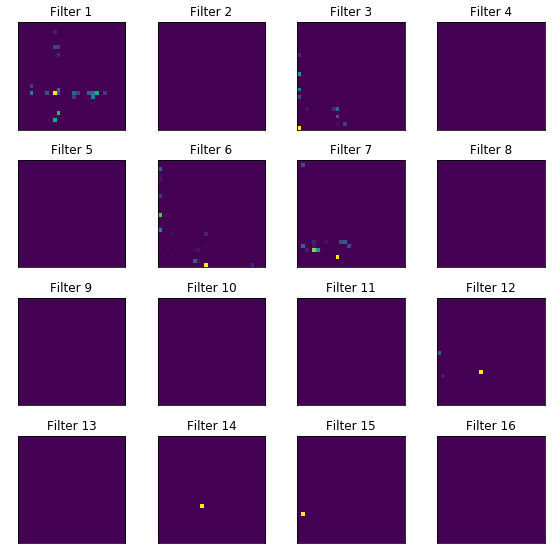}
        \caption{Participant ID: 1901, Song ID:3}
        \label{1901_3}
    \end{subfigure}
    \hspace{2mm}
    \begin{subfigure}[b]{0.45\textwidth}
        \includegraphics[width=\textwidth]{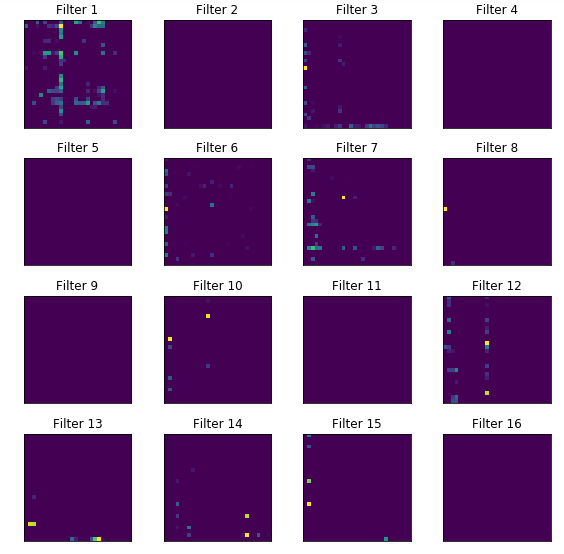}
        \caption{Participant ID: 1905, Song ID:3}
        \label{1905_3}
    \end{subfigure}
    \caption{Convolution layer 3 output}
    \label{conv3_out}
\end{figure}

We have visualized the intermediate CNN outputs. Fig.~\ref{conv3_out} shows the output of the 3\textsuperscript{rd} convolution layer of model 7. 
For the same song ID - 9, the participant 1901 and 1905 have learnt different features. 
Filter1, Filter7, Filter14, Filter15 shows different patterns in Fig.~\ref{1901_9}, \ref{1905_9}.
This supports our 2\textsuperscript{nd} hypothesis that the EEG patterns vary from person to person for a given song.
Similar analogy applies to Fig.~\ref{1901_3}, \ref{1905_3}.

\section{Conclusion}
In this paper, we proposed an approach to identify the song from brain activity, which is recorded when a person is listening to it. 
We worked on our data collected for 20 participants and 12 two minutes of songs having diverse tone, pitch, rhythm, and vocals.  
In particular, we were successfully able to classify songs from only 1 second long EEG response in frequency domain.
But, CNN model failed in time domain.
We developed a simple but yet efficient 3 layer deep learning model in the Keras framework. 
The results show that identifiable patterns are generated in the brain during music entrainment. 
We were able to detect them when each participant's EEG response considered in both train and test data. 
Our model performed poorly when some of the participants were completely excluded from the train data. 
This gives us insights about the different patterns created when different persons were listening to the same song. 
The possible reason could be people focus on a different tone, vocals during music entrainment, thereby reducing performance for cross-participant song identification task.
Thus, as future work, we aim at acquiring more data and look for other preprocessing methods, CNN architectures to improve accuracy for across participant data. 
However, results achieved in this paper are highly appreciable and provides an essential step towards an ambitious mind-reading goal. 

\section*{Author contributions}

DS developed the CNN model presented in this paper. KPM and DL were involved in experimental design, and discussion of results. BRS was involved in data collection and preprocessing EEG data. We would like to thank Ms. Esha Sharma for her contribution towards data collection. Data can be available on reasonable request to KPM.

%
%
%

\begin{thebibliography}{8}
\bibitem{b0} Brattico, E., Tervaniemi, M., Näätänen, R., Peretz, I.: Musical scale properties are automatically processed in the human auditory cortex. Brain research. 1117, 162–74 (2006).\doi{10.1016/j.brainres.2006.08.023}

\bibitem{b0_} Snyder, J., Large, E.: Gamma-Band Activity Reflects the Metric Structure of Rhythmic Tone Sequences. Brain research. Cognitive brain research. 24, 117–26 (2005).\doi{10.1016/j.cogbrainres.2004.12.014}

\bibitem{b1} Stober, S., Prätzlich, T., Müller, M.: Brain Beats: Tempo Extraction from EEG Data. In: ISMIR. pp. 276–282 (2016)

\bibitem{b2} Nozaradan, S.: Exploring how musical rhythm entrains brain activity with electroencephalogram frequency-tagging. Philosophical Transactions of the Royal Society B: Biological Sciences. 369, 20130393 (2014). \doi{10.1098/rstb.2013.0393}

\bibitem{b3}  Gang, N., Kaneshiro, B., Berger, J., Dmochowski, J.P.: Decoding Neurally Relevant Musical Features Using Canonical Correlation Analysis. In: ISMIR. pp. 131–138 (2017)

\bibitem{b4} Sanyal, S., Nag, S., Banerjee, A., Sengupta, R., Ghosh, D.: Music of brain and music on brain: a novel EEG sonification approach. Cogn Neurodyn. 13, 13–31 (2019). \doi{10.1007/s11571-018-9502-4}

\bibitem{b5} Yu, Y., Beuret, S., Zeng, D., Oyama, K.: Deep Learning of Human Perception in Audio Event Classification. 2018 IEEE International Symposium on Multimedia (ISM). (2018). \doi{10.1109/ISM.2018.00-11}

\bibitem{b6} Haynes, J.-D., Rees, G.: Decoding mental states from brain activity in humans. Nature Reviews Neuroscience. 7, 523–534 (2006)

\bibitem{b7} Bidelman, G., Powers, L.: Response properties of the human frequency-following response (FFR) to speech and non-speech sounds: level dependence, adaptation and phase-locking limits. International Journal of Audiology. 57, 665–672 (2018). \doi{10.1080/14992027.2018.1470338}

\bibitem{b8} Kavasidis, I., Palazzo, S., Spampinato, C., Giordano, D., Shah, M.: Brain2image: Converting brain signals into images. In: Proceedings of the 25th ACM international conference on Multimedia. pp. 1809–1817 (2017)

\bibitem{b9} Ntalampiras, S., Potamitis, I.: A Statistical Inference Framework for Understanding Music-Related Brain Activity. IEEE Journal of Selected Topics in Signal Processing. 13, 275–284 (2019). \doi{10.1109/JSTSP.2019.2905431}

\bibitem{b10} Stober, S., Sternin, A., Owen, A.M., Grahn, J.A.: Towards Music Imagery Information Retrieval: Introducing the OpenMIIR Dataset of EEG Recordings from Music Perception and Imagination. In: ISMIR. pp. 763–769 (2015)

\bibitem{b11} Foster, C., Dharmaretnam, D., Xu, H., Fyshe, A., Tzanetakis, G.: Decoding Music in the Human Brain Using EEG Data. In: 2018 IEEE 20th International Workshop on Multimedia Signal Processing (MMSP). pp. 1–6 (2018). \doi{10.1109/MMSP.2018.8547051}

\bibitem{b13} Stober, S., Cameron, D.J., Grahn, J.A.: Using Convolutional Neural Networks to Recognize Rhythm Stimuli from Electroencephalography Recordings. In: Ghahramani, Z., Welling, M., Cortes, C., Lawrence, N.D., and Weinberger, K.Q. (eds.) Advances in Neural Information Processing Systems 27. pp. 1449–1457. Curran Associates, Inc. (2014)

\bibitem{vgg}  Long, J., Shelhamer, E., Darrell, T.: Fully Convolutional Networks for Semantic Segmentation. Presented at the Proceedings of the IEEE Conference on Computer Vision and Pattern Recognition (2015)

\bibitem{densenet} Iandola, F., Moskewicz, M., Karayev, S., Girshick, R., Darrell, T., Keutzer, K.: Densenet: Implementing efficient convnet descriptor pyramids. arXiv preprint arXiv:1404.1869. (2014)

\bibitem{relu} Nair, V., Hinton, G.E.: Rectified linear units improve restricted boltzmann machines. In: Proceedings of the 27th international conference on machine learning (ICML-10). pp. 807–814 (2010)

\bibitem{keras} F. Chollet et al., “Keras,” \url{https://github.com/fchollet/keras}, 2015

\end{thebibliography}
%

\end{document}